\documentclass[sigconf, nonacm, article]{acmart}

\PassOptionsToPackage{table}{xcolor} 
\usepackage{xcolor}

\usepackage{graphicx}
\usepackage{multirow} 
\usepackage{rotating} 
\usepackage{float} 
\usepackage[inkscapelatex=false]{svg}

\usepackage{enumitem}
\usepackage{array}
\usepackage{geometry}
\usepackage{tabularx}

\AtBeginDocument{%
  }

\setcopyright{acmlicensed}
\copyrightyear{2025}
\acmYear{2025}
\acmDOI{XXXXXXX.XXXXXXX}

\acmConference[Conference acronym 'XX]{Make sure to enter the correct
  conference title from your rights confirmation emai}{June 03--05,
  2018}{Woodstock, NY}
\acmISBN{978-1-4503-XXXX-X/18/06}

\begin{document}



\title{Intelligent Tutors Beyond K-12: An Observational Study of Adult Learner Engagement and Academic Impact}

\author{Adit Gupta}
\affiliation{%
  \institution{Drexel University}
  \streetaddress{3230 Market Street}
  \city{Philadelphia} 
  \state{PA}
  \country{USA}}

\author{Christopher MacLellan}
\affiliation{%
  \institution{Georgia Institute of Technology}
  \streetaddress{North Avenue}
  \city{Atlanta}
  \state{GA}
  \country{USA}}

\begin{abstract}
Intelligent tutors have proven to be effective in K-12 education, though their impact on adult learners---especially as a supplementary resource---remains underexplored. Understanding how adults voluntarily engage with educational technologies can inform the design of tools that support skill re-learning and enhancement. More critically, it helps determine whether tutoring systems, which are typically built for K-12 learners, can also support adult populations. This study examines the adoption, usage patterns, and effectiveness of a novel tutoring system, Apprentice Tutors, among adult learners at a state technical college. We analyze three types of data including, user demographics, grades, and tutor interactions, to assess whether voluntary tutor usage translates into measurable learning gains. Our findings reveal key temporal patterns in tutor engagement and provide evidence of learning within tutors, as determined through skill improvement in knowledge components across tutors. We also found evidence that this learning transferred outside the  tutor, as observed through higher course assessment scores following tutor usage. These results suggest that intelligent tutors are a viable tool for adult learners, warranting further research into their long-term impact on this population.
\end{abstract}


\maketitle

\section{Introduction}
Over the past several decades, intelligent tutoring systems (ITS) have seen significant improvements and gained commercial adoption. Examples of platforms that have seen global adoption include Duolingo and Khan Academy, which attract millions of voluntary users \cite{vesselinov2012duolingo, murphy2014research}. Duolingo, for example, has reported more than 500 million registered users learning languages through its platform \cite{vesselinov2012duolingo}. These platforms are particularly compelling because they cater to a diverse set of users who choose to engage with and repeatedly use these systems voluntarily. Moreover, both Duolingo and Khan Academy leverage psychological principles, such as gamification and spaced repetition, to enhance the retention and motivation of learners \cite{deterding2011gamification, cepeda2006distributed}.

In addition to commercialization, intelligent tutors have been built and studied in research contexts since the early 1970s \cite{carbonell1970ai, maclellan2022domain, Clancey_1986, pane2014effectiveness, vanlehn2011, wijekumar2012large}. Early research on tutoring systems demonstrated that tutor use leads to improved learning outcomes \cite{corbett1995knowledge, koedinger1997intelligent}. For example, Corbett and Anderson's knowledge tracing model showed that intelligent tutors could effectively model student learning and adapt instruction accordingly, resulting in enhanced procedural knowledge acquisition \cite{corbett1995knowledge}. Researchers have shown that these learning gains transfer beyond the tutor's environment, including within physical classroom settings \cite{pane2014effectiveness, koedinger1997intelligent}. Pane et al. \cite{pane2014effectiveness} conducted a randomized large-scale study on Cognitive Tutor Algebra and found that students who used tutors showed significant improvements in standardized test scores compared to those receiving traditional instruction. This suggests that intelligent tutors not only facilitate immediate learning, but also contribute to long-term academic achievements.

Much of the previous research on tutoring systems has focused on K-12 students in controlled experiments where tutor usage is strictly prescribed. For example, in controlled studies by Razzaq and Heffernan \cite{razzaq2010tutor} and Mitrovic and Ohlsson \cite{mitrovic1999evaluation}, students were required to use an intelligent tutoring system with specific pretests and posttests to evaluate the effect of tutoring usage and the corresponding learning gains. Studies within this research direction have consistently found that mandatory tutor use leads to measurable improvements in student understanding and skills. However, there remains a gap in the literature that examines the use of tutors beyond K-12 populations, particularly in contexts where involvement is entirely voluntary. 

To address this gap, our work is focused on adult learners that uses tutors voluntarily. Adult learners are inherently different from K-12 learners in many aspects, including their intrinsic motivations and schedules \cite{knowles2015adult, cercone2008characteristics}. This user group may be balancing full-time employment and additional education as competing priorities. Understanding how adult learners adopt and use tutoring systems in these contexts can help us design intelligent tutors tailored to their unique needs across a wide range of domains, especially for those seeking to re-skill. The primary aim of this work was to investigate how voluntary usage of intelligent tutors affects adult learners by exploring tutor user demographics and assessing how such usage translates into improved classroom learning outcomes. As such, this work is guided by two core research questions:
\begin{enumerate}
    \item How does the voluntary use of intelligent tutors affect adult learners' outcomes?
    \item What unique characteristics and demographic factors define adult learners who voluntarily choose to use intelligent tutors?
\end{enumerate}

To explore these questions, we present an observational multi-method study, examining a two-year deployment of the Apprentice Tutors system at a 2-year state technical college. We examined metadata from this tutor deployment with the aim of understanding usage patterns. This work presents three primary contributions to the field of intelligent tutors and educational data mining:
\begin{enumerate}
    \item Preliminary evidence suggests that Apprentice Tutors are effective as supplementary resources for adult learners. 
    \item A unique adult learner dataset that includes information about tutors usage, demographics, and course outcomes, thereby focusing on an important and historically understudied user population.
    \item Observations from this study are related to previously established adult learner design recommendations that were presented in \cite{gupta2024intelligent}, with the aim of building continuity between adult learner needs and tutor usage patterns observed in the data.
\end{enumerate}

This work advances educational technology by offering empirical insights into how adult learners engage with pedagogical tools. Ultimately, these insights can empower this population, which includes professionals seeking growth, leading to broad educational and socioeconomic benefits.

\section{Related Works}
Previous research has shown that intelligent tutors can significantly improve learning outcomes, particularly in K-12 settings, where their use is often mandatory. For example, Koedinger et al. \cite{koedinger1997intelligent} found that high school students using Cognitive Tutor for Algebra I outperformed their peers receiving traditional instruction on standardized tests. Similarly, a meta-analysis by Ma et al. \cite{ma2014intelligent} concluded that students using intelligent tutors exhibited substantial gains in academic achievement compared to those receiving conventional classroom instruction. VanLehn's analysis \cite{vanlehn2011relative} suggests that intelligent tutors are nearly as effective as human tutors and sometimes more effective than traditional classroom instruction. 

Although intelligent tutors have been extensively studied in K-12 contexts, adult learners represent a distinct demographic with unique learning characteristics. Adult learners differ from younger students in several key aspects, including greater self-direction, accumulated life experiences, and a focus on practical application \cite{knowles1984adult,  knowles2015adult}. According to the principles of andragogy, adults are intrinsically motivated and prefer learning that is relevant to their personal and professional lives \cite{merriam2001andragogy, merriam2014adult}. Adult learners tend to be self-directed learners who value autonomy in the learning process \cite{garrison1997self, hiemstra2012reframing}. In addition, they often balance multiple responsibilities, such as work and family, which can affect their engagement with educational technologies \cite{cercone2008characteristics, brookfield2013powerful}.

Despite a consistent track record of improving student learning, tutors have not been widely adopted by adult learners across non-traditional educational environments, such as workplace training programs, online education platforms, and continuing professional development courses. However, there has been work exploring the use of tutoring systems to support adult learners in specialized contexts. For example, the GIFT (Generalized Intelligent  Framework for Tutoring) framework was developed to support tutor creation, with a specific consideration to the unique needs of military personnel, focusing on skills acquisition and decision-making under pressure \cite{Sinatra2014}. Furthermore, Lippert et al. \cite{lippert2020multiple} investigated intelligent tutors for adults, emphasizing adaptability and real-time feedback to enhance operational readiness and training effectiveness. Nevertheless, this prior work primarily targets adult learners in high-stakes environments and does not always directly translate to real-world classroom settings involving adult students pursuing non-military educational goals.

Kizilcec and Schneider \cite{kizilcec2015motivation} explored how adults engage with adaptive learning technologies in massive open online courses (MOOCs) and found that personalized feedback can enhance engagement and persistence. However, a gap in this work is the lack of  insight into how users engage in MOOCs, from a session adoption and retention standpoint. Further, this work does not fully address the influence of motivation and self-regulation on their learning behaviors and engagement. 


When intelligent tutor usage is optional, adults' adoption and engagement can vary widely. The Technology Acceptance Model (TAM) proposed by Davis \cite{davis1989perceived} indicates that perceived usefulness and ease of use are critical factors that influence technology adoption. Barriers such as time constraints, lack of technical skills, and low self-efficacy can hinder voluntary usage \cite{sun2006role,cercone2008characteristics}. Beyond such external factors, demographic factors such as age, gender, and ethnicity can impact the adoption and effectiveness of intelligent tutors among adult learners. Zawacki-Richter and Latchem \cite{zawacki2014educational} observed that older adults may be less inclined to use educational technologies due to lower digital literacy. Gender differences have also been reported; Ong and Lai \cite{ong2006gender} found that women may experience greater computer anxiety, affecting their engagement levels. 

Understanding these demographic influences is key to designing tutors that work well for diverse adult populations. We see an opportunity to extend current intelligent tutor research by combining critical data on tutor adoption, user demographics, and learning outcomes into a single study. While previous work often treats these data separately, integrating them provides a clearer picture of how these technologies impact users. This gap highlights the need for observational studies that analyze real-world usage data to understand how users voluntarily engage with educational technologies. By examining both usage patterns and academic performance, this research offers a more nuanced view of how intelligent tutors support adult learners in supplemental educational settings.

\begin{figure*}
    \centering
    \includegraphics[width=1.0\textwidth]{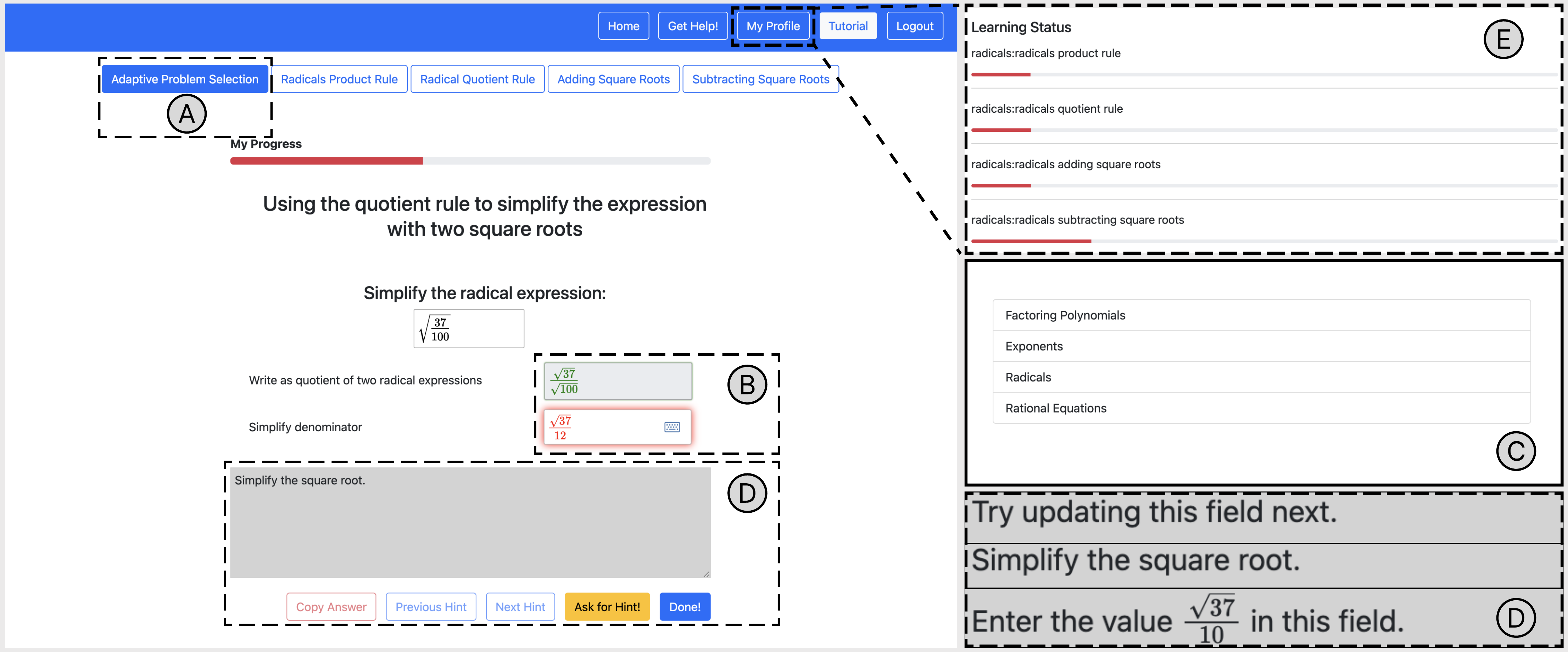} 
    \caption{User interface of Apprentice Tutors platform with key features: (a) penalization through adaptive problem selection (b) real-time correctness feedback (c) four available tutors (d) hint box and multi-layer hints (e) user profile screen with progress bars corresponding to knowledge components.}
    \label{fig:apprentice}
\end{figure*}

\section{Apprentice Tutors}
This study uses metadata from a novel tutoring system for adult learners, Apprentice Tutors platform \cite{gupta2024intelligent}. Developed through extensive cognitive task analysis, the platform offers ten tutors for College Algebra topics, including radicals, exponents, factoring polynomials, and rational equations. A complete list of supported topics is provided in Table \ref{syllabus}. Each tutor consisted of multiple problem types with expert models, interfaces, and problem generators, enabling real-time correctness feedback and multi-layered hints. The platform leverages Learning Tools Interoperability (LTI) standards to facilitate seamless deployment in learning management systems like Blackboard and Canvas.

\begin{figure}[htp]
    \centering
    \includegraphics[width=1\columnwidth]{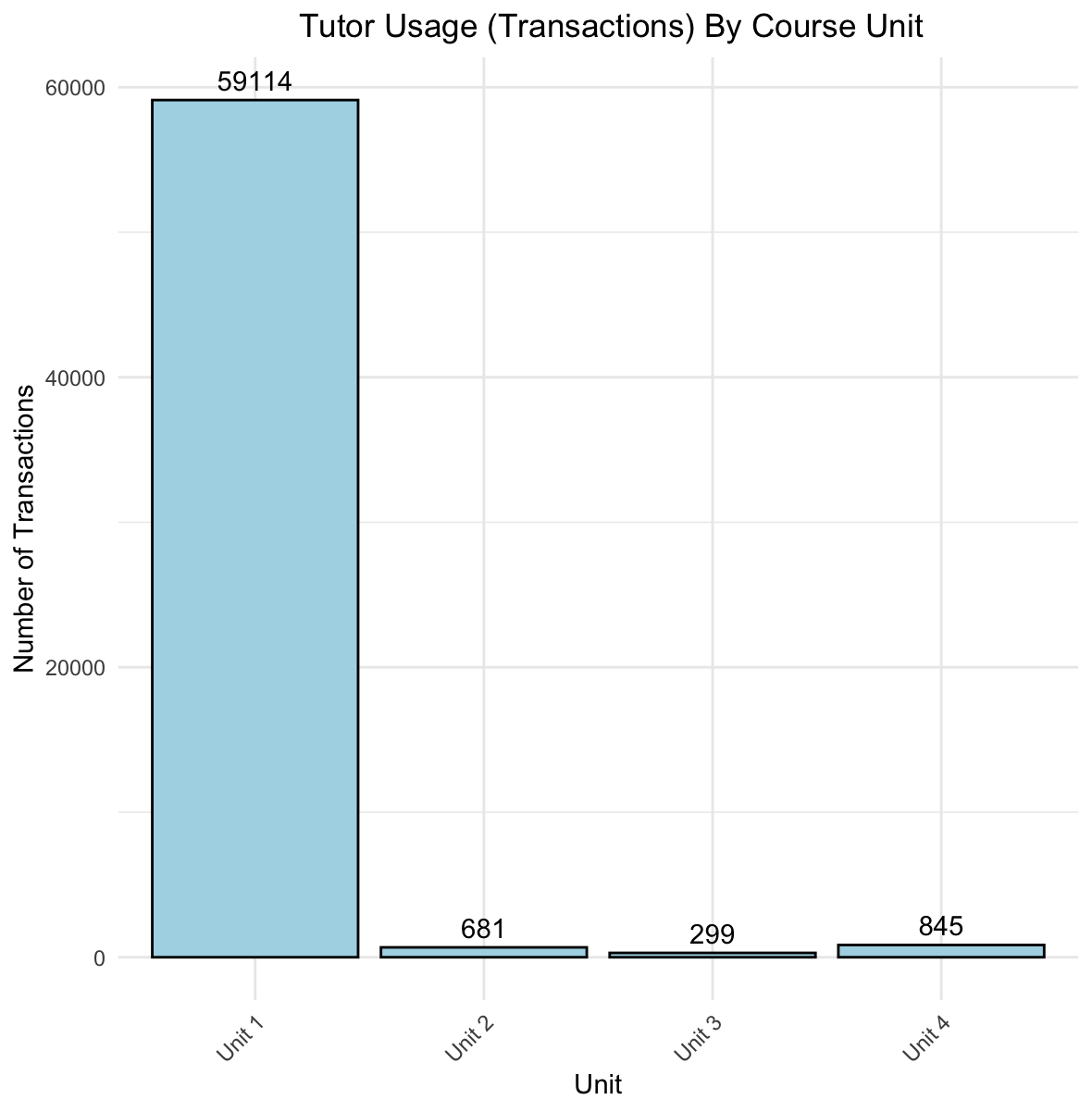}
    \caption{Tutor usage by course unit in College Algebra. The x-axis represents the different units, and the y-axis shows the number of transactions recorded for each unit. Unit 1 has significantly higher tutor usage compared to other units, suggesting a higher tutor coverage in this area.}
    \label{fig:unit_usage}
\end{figure}

Figure \ref{fig:apprentice} shows the user interface of the Apprentice Tutors platform, which incorporates several key features designed to adapt to student needs. The adaptive problem selection utilizes Bayesian Knowledge Tracing (BKT) \cite{corbett1995knowledge} to personalize the learning experience by selecting problems that target unmastered skills. The interface provides real-time correctness feedback, immediately indicating whether the student's input is correct or incorrect. Multi-layered hints are available through the hint box to guide students when needed. The user profile screen tracks progress on different knowledge components, allowing students to monitor their mastery of the material. By closely aligning the tutors with the course curriculum and generating randomized problems, the platform aims to enhance student engagement and academic performance. The combination of adaptive problem selection, real-time feedback, and contextual hints enhances students' engagement and facilitates mastery of the course material.

We extended the Apprentice Tutors, based on the thematic analysis \cite{Braun2006} presented in \cite{gupta2024intelligent}. First, tutorial videos were created for each problem to demonstrate how students could engage effectively with each tutor. Second, a bug submission and Q\&A chat feature was integrated just before the focus group, allowing users to report bugs or issues and directly ask questions to team members when they encountered difficulties. Third, each tutor interface was modified to include direct links to relevant learning materials from the OpenStax \textit{Intermediate Algebra, 2e} textbook \cite{openstax_intermediate_algebra_2e} and to relevant YouTube resources, recognizing that many students turn to YouTube for additional support. Fourth, more detailed and directed hint messages were incorporated. Finally, we developed several additional tutors, see Figure \ref{syllabus} for a complete list.

\begin{table}[htbp]
\small
\centering
\begin{tabular}{|>{\raggedright\arraybackslash}p{6cm}|c|}
\hline
\textbf{Course Topics} & \textbf{Coverage} \\
\hline \textbf{Unit 1: Algebraic Expressions} & \\ \hline Exponents & \textbf{Yes} \\ Radicals and Rational Expressions & \textbf{Yes} \\ Polynomials & No \\ Factoring Polynomials & \textbf{Yes} \\ Rational Expressions & \textbf{Yes} \\ The Rectangular Coordinate Systems and Graphs & No \\ \hline \textbf{Unit 2: Equations and Linear Inequalities} & \\ \hline Linear and Rational Equations in One Variable & \textbf{Yes} \\ Models and Applications of Linear Equations & No \\ Systems of Linear Equations: Two Variables & No \\ Complex Numbers & No \\ Quadratic Equations & \textbf{Yes} \\ \hline \textbf{Unit 3: Functions and More Inequalities} & \\ \hline Introduction to Sets & No \\ Linear Inequalities & No \\ Quadratic Inequalities & No \\ Rational Inequalities & No \\ Functions and Function Notation & No \\ Domain and Range & No \\ Linear Functions & \textbf{Yes} \\ Quadratic Functions & \textbf{Yes} \\ \hline \textbf{Unit 4: Exponential and Logarithmic Functions} & \\ \hline Exponential Functions and their Graphs & No \\ Logarithmic Functions and their Graphs & \textbf{Yes} \\ Logarithmic Properties & No \\ Exponential and Logarithmic Equations & \textbf{Yes} \\ \hline \end{tabular}
\caption{College Algebra Course Topics with Apprentice Tutor Coverage Highlighted}
\label{syllabus}
\end{table}

The Apprentice Tutors platform currently supports a College Algebra course, which consists of four academic units, as detailed in Table \ref{syllabus}. The tutors were developed by modeling the course lessons identified in OpenStax's \textit{Intermediate Algebra, 2e}, specifically the Intermediate Algebra 2 course section \cite{openstax_intermediate_algebra_2e}. Unit 1 comprises six topics, four of which are covered by tutors. Unit 2 includes five topics, with two covered by tutors. Unit 3 consists of eight topics, two of which are supported by the tutors, and Unit 4 has four topics, two of which are covered by the tutors. Although the tutors provide support for topics across all units, Unit 1 has the highest overlap with the available tutors. Figure \ref{fig:unit_usage} shows the distribution of logged tutor transactions between the four units. The data show that tutor usage is highest during the first unit, exceeding the combined usage of the other units. Given the extensive topic coverage and the increased usage of tutors in Unit 1, this study will focus on this particular unit.

\section{Methodology}
\subsection{Study Design}
This is an observational, quasi-experimental study to investigate the adoption, usage patterns, and effectiveness of the Apprentice Tutors platform among adult learners. This study design aligns with established methodologies when investigating the impact of intelligent tutors, to capture usage behavior, and does so without manipulating the learning environment \cite{koedinger1997intelligent, aleven2004towards}. This section provides more context on the sources of data, participation recruitment, and analysis methods used. 

\subsection{Participants and Learning Context}
The Apprentice Tutors were offered to students enrolled in the College Algebra course at a 2-year state technical college. This institution offers a unique learning environment geared towards adult learners who want to further their education by going back to school. Many of the students here are adult learners who work full-time and take night classes or online classes in addition to their full-time work. To offer flexibility, this college offers multiple course formats including in-person, hybrid, and fully online. The College Algebra course is structured into four units, with the Apprentice Tutors platform offering support for topics in each unit (see Table \ref{syllabus}). The student population comprised adult learners with diverse backgrounds, including varying ages, genders, and ethnicities, reflecting the heterogeneity of the adult learner demographic. 

\subsection{Data Collection}
\subsubsection{Tutor Deployment}
The Apprentice Tutors platform was integrated into existing learning management systems (Blackboard and Canvas) using Learning Tools Interoperability (LTI) standards. Tutors were made available as optional supplemental resources, allowing students to engage with the system voluntarily. This deployment strategy mirrors real-world scenarios in which adult learners choose to use available educational resources based on their needs and schedules. 
 
\subsubsection{Grades and Demographics}
We collected academic performance data from the institution. This data provided detailed information on each student enrolled in the College Algebra course. This data included performance metrics and results of the course for specific assessments and unit examinations for each individual in the college algebra course, in all sections. In addition, we also obtained demographic data that included each student's ethnicity, age, and gender. 

A significant challenge in this study was the integration of various data sets while preserving user anonymity. To achieve this, we used the unique (anonymous) Learning Tools Interoperability (LTI) identification string to securely link data from tutors, grades, and demographics. Learning Tools Interoperability (LTI) is a standard protocol that enables seamless integration of various educational tools and platforms \cite{LTI_standard_citation}.

\subsection{Data Analysis}
A combination of statistical methods was used to analyze the data:
\begin{itemize}
\item \textbf{Descriptive Statistics:} Summarized demographic distributions, usage patterns and performance metrics to provide an overview of the dataset. 
\item \textbf{Usage Analysis:} Applied to identify patterns in usage behaviors and group students with similar interaction profiles. This helps in understanding how different user groups engage with the ITS.
\item \textbf{Learning Curve Analysis:} Used to model knowledge component mastery over time, identifying trends in skill acquisition and the effectiveness of tutors in promoting learning.
\item \textbf{Regression Analysis:} Employed to examine the relationship between tutor usage levels and learning outcomes. This approach helps identify the impact of tutor engagement on academic performance. 
\end{itemize}
By integrating these analytical methods, this work aims to provide a comprehensive understanding of how adult learners interact with tutors and the resulting impact on their learning outcomes.

\section{Results and Analysis}
\subsubsection{Demographic Patterns in Tutor Usage}

\begin{figure}[htp]
    \centering
    \includegraphics[width=\columnwidth]{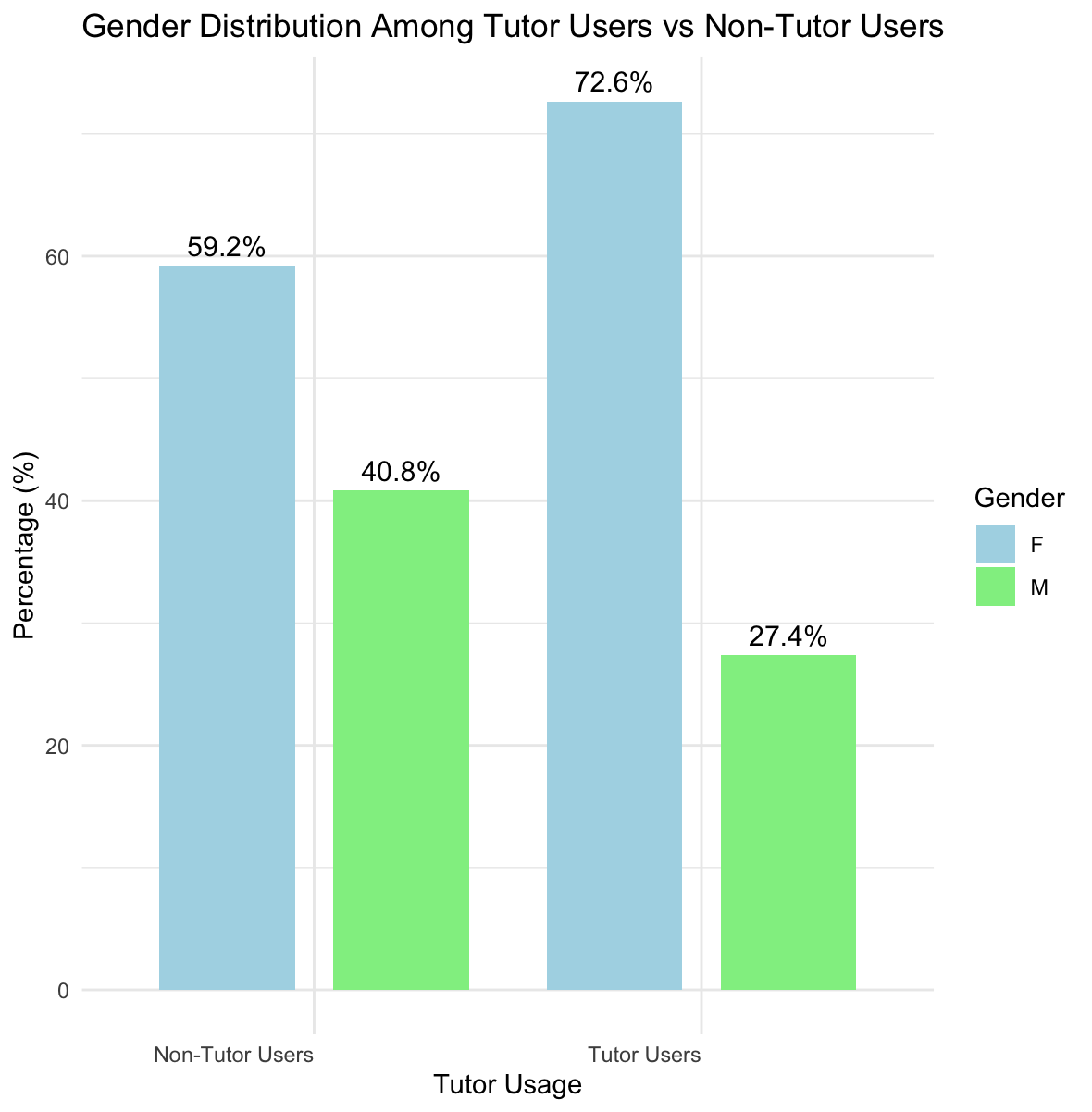}
    \caption{Demographic Data of Apprentice Tutors: Gender distribution comparison between tutor and non-tutor users.}
    \label{fig:demographics_gender}
\end{figure}

\begin{figure}[htp]
    \centering
    \includegraphics[width=\columnwidth]{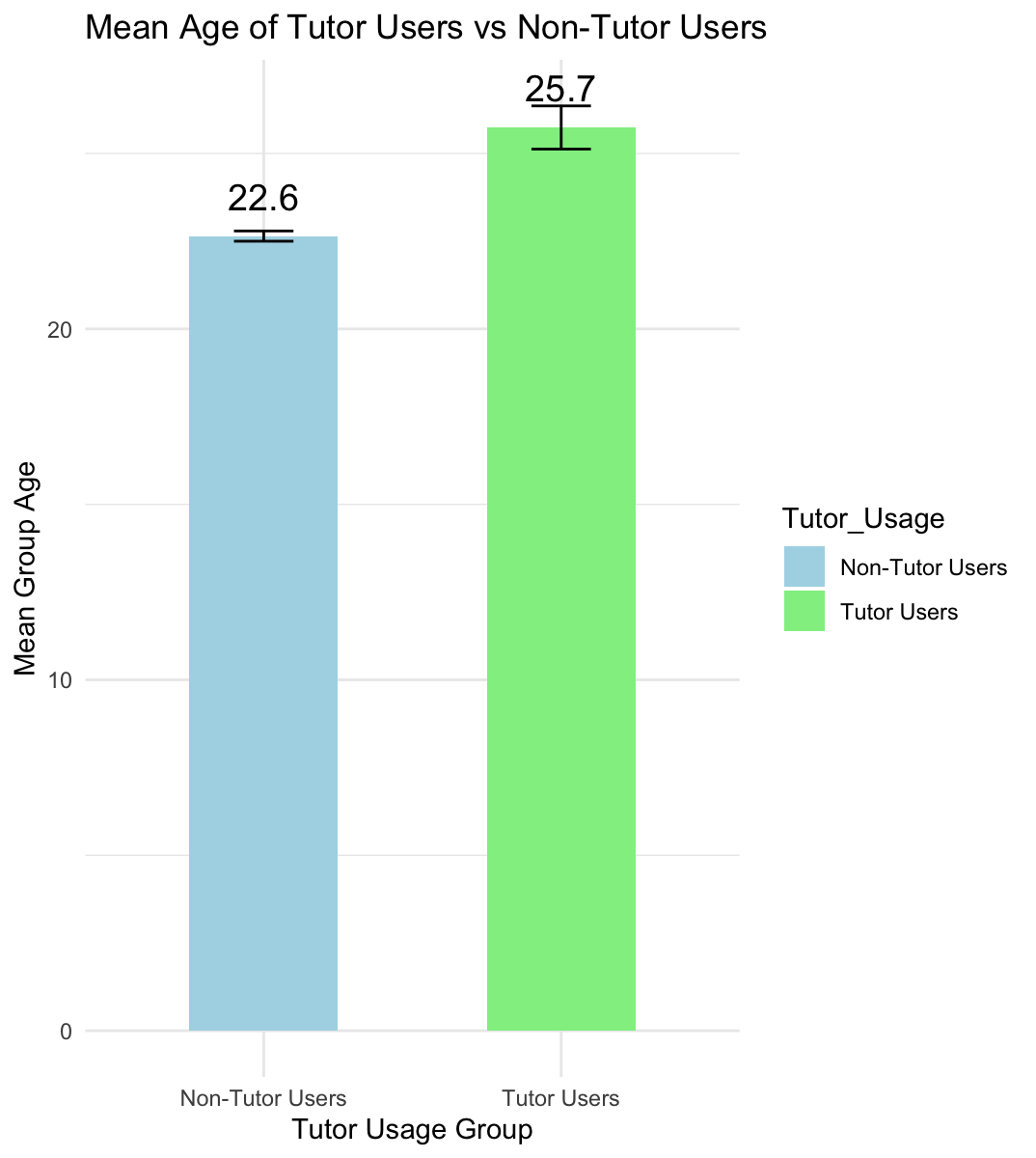}
    \caption{Demographic Data of Apprentice Tutors: Age distribution comparison between tutor and non-tutor users.}
    \label{fig:demographics_age}
\end{figure}

\begin{figure}[htp]
    \centering
    \includegraphics[width=\columnwidth]{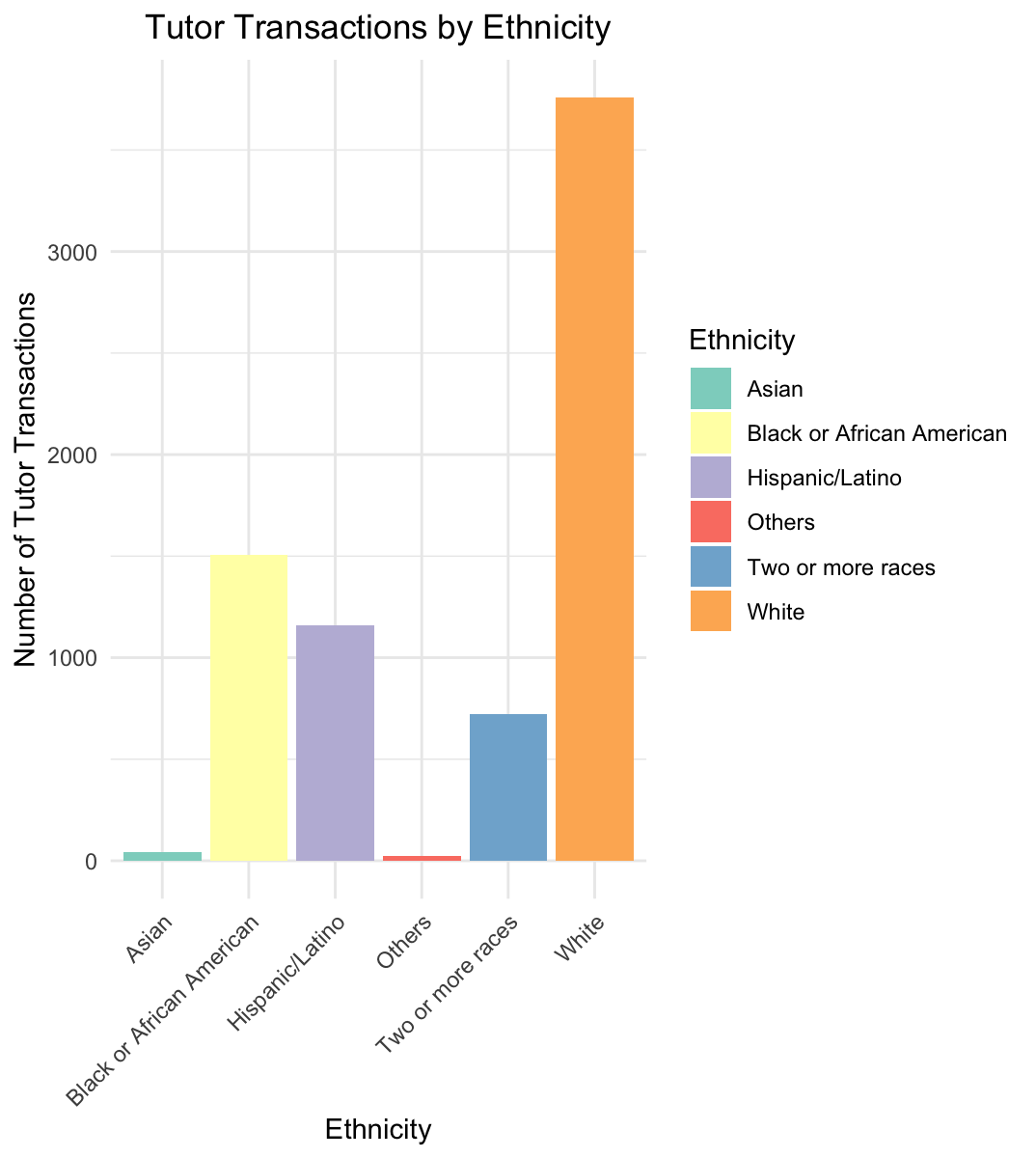}
    \caption{Demographic Data of Apprentice Tutors: Ethnicity distribution comparison between tutor and non-tutor users.}
    \label{fig:demographics_ethnicity}
\end{figure}

In this section, we share several demographic trends in the adoption and usage of Apprentice Tutors. In the specified time period, 3,510 unique students were recorded, of which 2,990 (85.18\% of 3,510) did not use the tutors, and 520 (14.81\% of 3,510)  students voluntarily used the tutors at least once. Of those 520 students, 214 (41.15\% of 520) unique students used the platform at least once during Unit 1 of the College Algebra course. 

Starting with gender observations, Figure \ref{fig:demographics_gender} illustrates the gender distribution among tutor users compared to non-tutor users. Among non-tutor users, the gender distribution is relatively balanced, with 59.2\% identifying as female and 40.8\% as male.\footnote{We acknowledge that gender is not binary; however these were the only two categories in the data reported by the technical college.} However, among tutor users, there is a significant increase in the proportion of females, which rose to 72.6\%, while the proportion of males decreased to 27.4\%. As shown in Figure \ref{fig:demographics_age}, Tutor users were, on average, approximately three years older than non-Tutor users (25.7 vs. 22.6 years). In terms of ethnicity, Figure \ref{fig:demographics_ethnicity} illustrates the amount of tutor usage across various self-reported ethnic groups. The largest user segments included white and black or african-american users, followed by hispanic/latino users as the third largest user segment. 

\begin{figure}
    \centering
    \includegraphics[width=1\columnwidth]{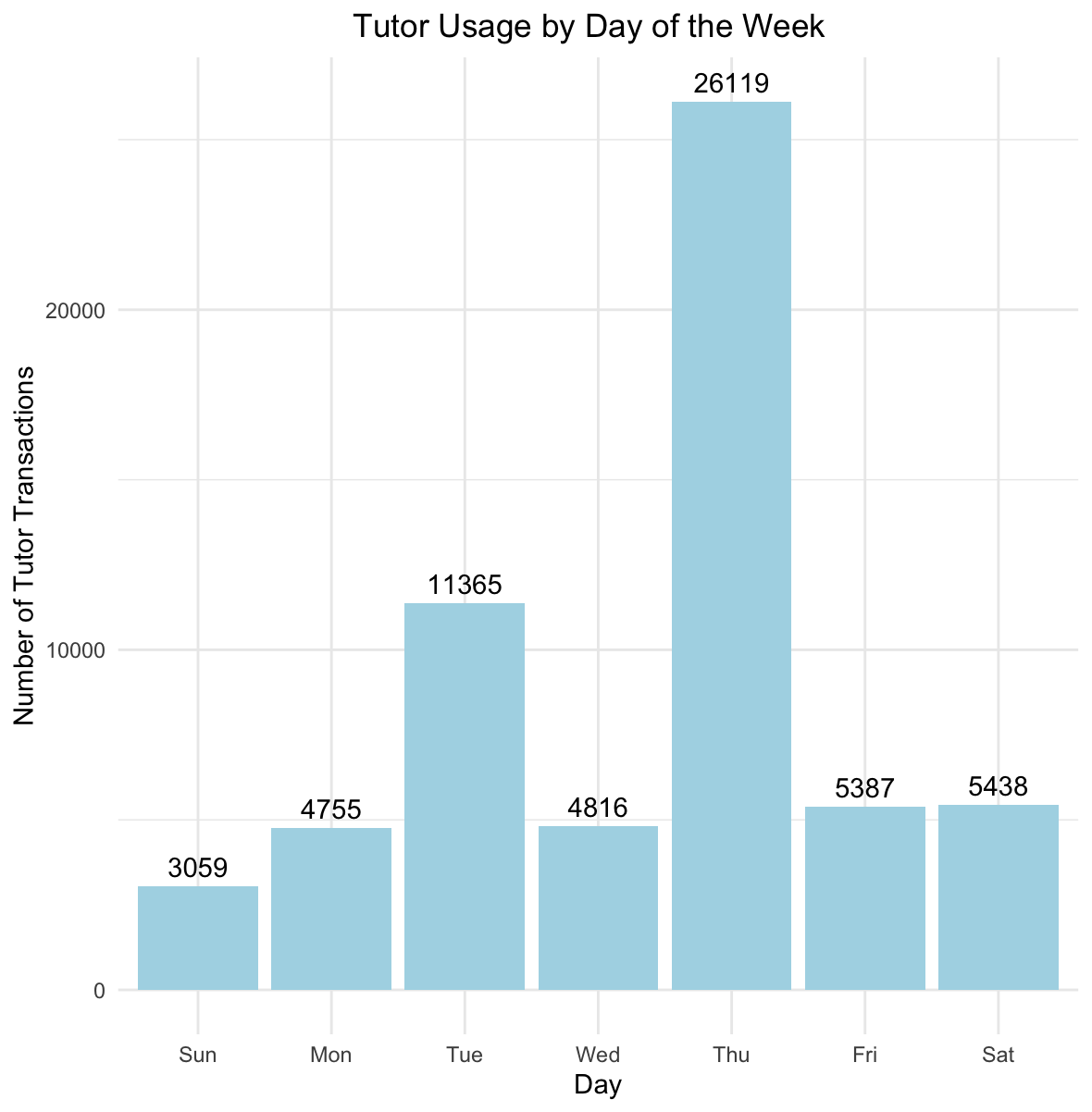}
    \caption{Tutor Usage by Day of the Week. The number of tutor transactions varies by day, with the highest usage on Thursday.}
    \label{fig:tutor_usage_day}
\end{figure}

\begin{figure}
    \centering
    \includegraphics[width=1.0\columnwidth]{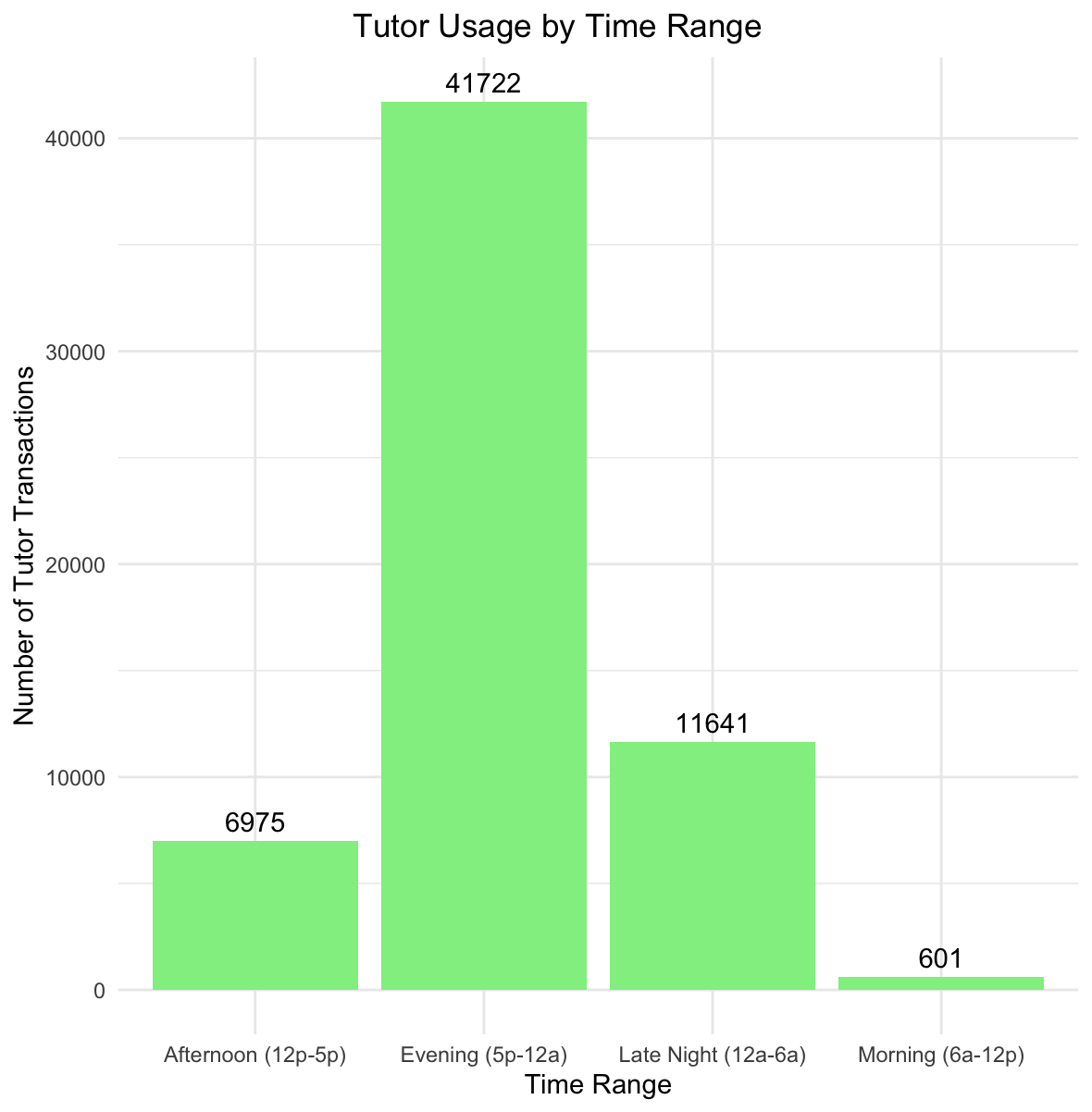}
    \caption{Tutor Usage by Time Range. The distribution of tutor usage across time ranges shows the highest usage during the evening (5 p.m. to 12 a.m.), followed by late night (12 a.m. to 6 a.m.)}
    \label{fig:tutor_usage_time}
\end{figure}

Beyond demographic characteristics, temporal usage patterns provide additional insights into when adult learners engage with the tutors. This analysis was motivated by findings from \cite{gupta2024intelligent}, in which learners expressed a desire to access tutors at any time, highlighting the need for continuous availability to support diverse learning schedules. Figure \ref{fig:tutor_usage_day} indicates that Thursdays saw the highest tutor usage, with usage also peaking during the evenings (5 p.m. to 12 a.m.) and late nights (12 a.m. to 6 a.m.), as shown in Figure \ref{fig:tutor_usage_time}. These temporal patterns suggest that adult learners are using tutoring resources outside of traditional work hours, aligning with the goal of providing flexible and accessible educational support. Understanding these usage trends is crucial for optimizing tutor availability and ensuring that the tutoring system meets the needs of adult learners.

\subsubsection{Learning Outcomes and Progress in Tutor Users}

\begin{figure}[t!]
    \centering
    \includegraphics[width=1.0\columnwidth]{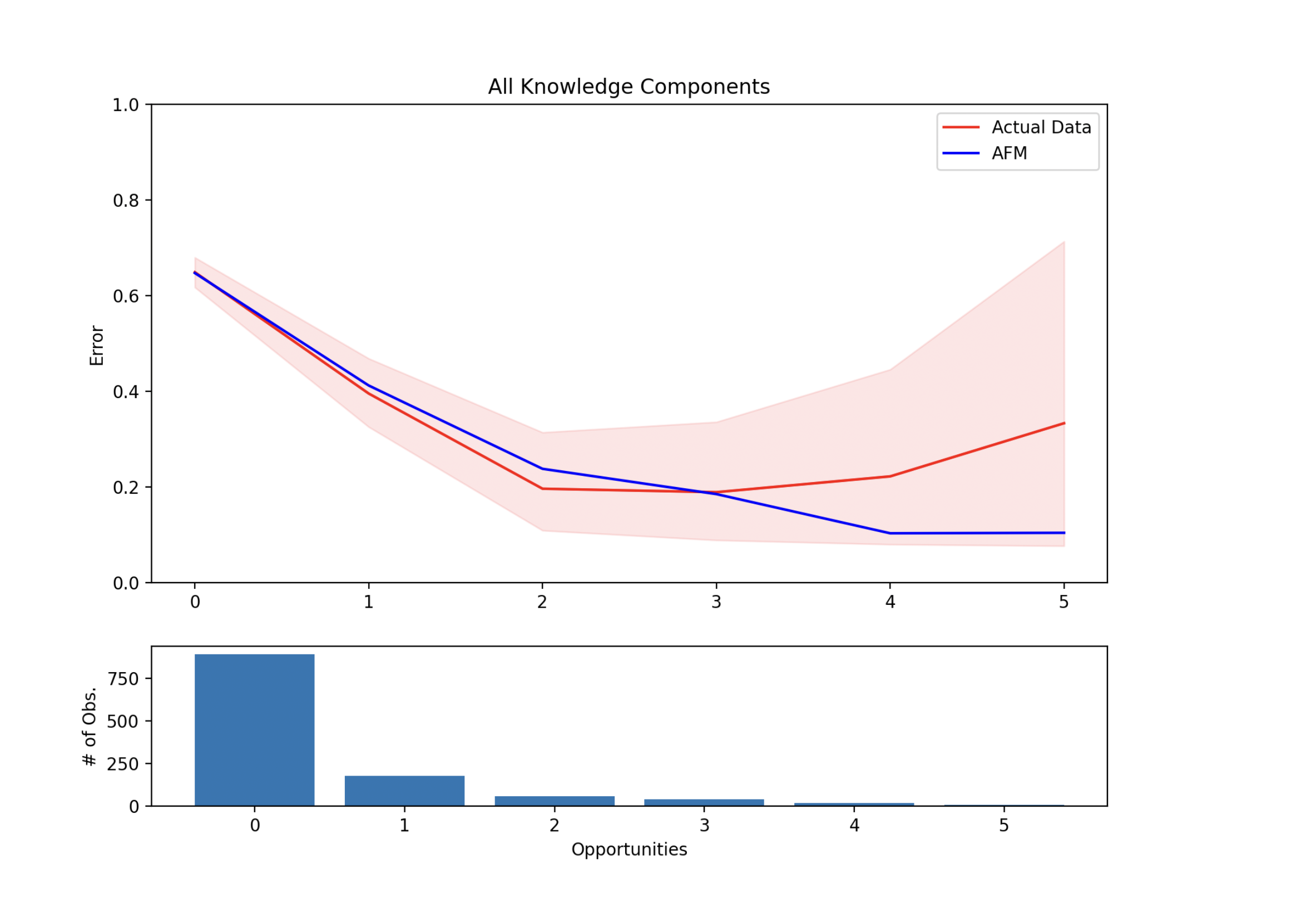}
    \caption{Decrease in student error rates with practice ($p < 0.001$). The x-axis represents the number of skill practice opportunities, while the y-axis represents the predicted error rate. The red shaded area indicates the 95\% confidence interval around the average.}
    \label{fig:learning_curve}
\end{figure}

To assess the impact of the tutors on skill acquisition, we analyzed student performance trends over time within the Apprentice Tutors. Figure \ref{fig:learning_curve} illustrates a learning curve that demonstrates that student error rates decreased with increasing practice opportunities. This figure was generated using \textbf{pyAFM}, a Python library to model student learning data based on the Additive Factors Model (AFM), a logistic regression model widely used in learning curves research \cite{macllellan2016impact}. \textbf{pyAFM} estimates key learning parameters such as initial levels of knowledge of students, the difficulty of the skills being taught, and the rate at which these skills are acquired.

As shown in Figure \ref{fig:learning_curve}, the students initially had an average error rate of approximately 65\% per skill on their first attempt. If students requested a hint or entered the correct answer after seeking assistance, these responses were still marked as ``incorrect.'' However, with continued practice, the error rates decreased significantly, highlighting the effectiveness of tutors in facilitating learning and improvement.




\subsubsection{Effect of Tutor Usage on Class Performance}

\begin{figure}[t!]
    \centering
    \includegraphics[width=1\columnwidth]{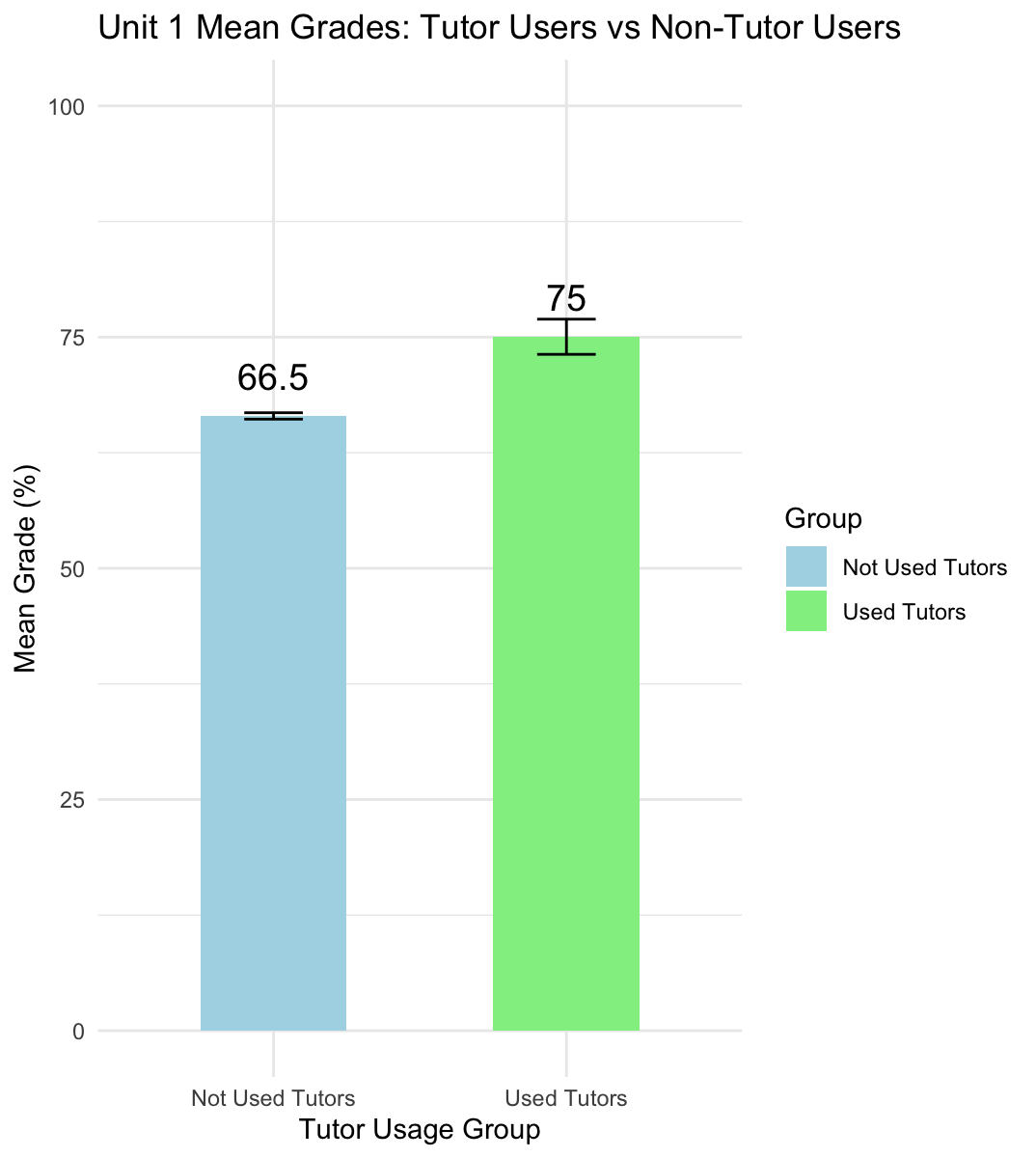}
    \caption{Mean grades on Unit 1 assessment: Comparison between Tutor Users and Non-Tutor Users. This bar chart shows the averages with 95\% confidence intervals.}
    \label{fig:unit_1_grades_comparison}
\end{figure}

Beyond in-tutor learning, we also examined whether tutor usage correlated with improved classroom performance. Figure \ref{fig:unit_1_grades_comparison} shows that tutor users achieved slightly higher mean grades in Unit 1 assessments compared to non-tutor users. We observed a total of 3,510 students, of which 520 students accessed tutors throughout the deployment. Of these, 214 students had at least one interaction with the unit 1 tutors. The decision to focus solely on Unit 1 was driven by its extensive tutor coverage and the highest level of tutor engagement among all course units. As detailed in Table \ref{syllabus}, Unit 1 comprises six topics, four of which are supported by the Apprentice Tutors. 

\begin{figure}[t!]
    \centering
    \includegraphics[width=1.0\columnwidth]{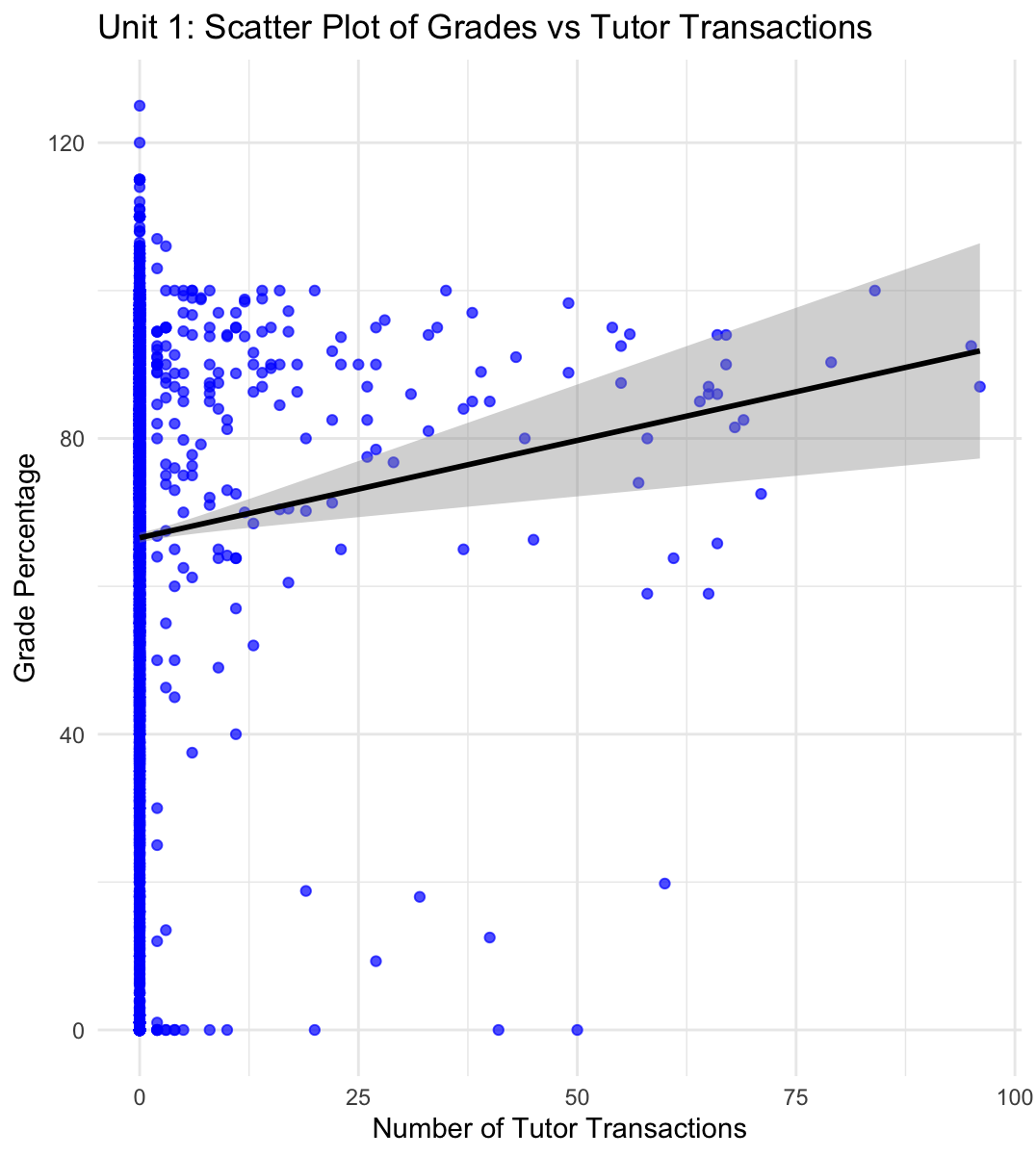}
    \caption{Unit 1: Scatter Plot of Grades vs Tutor Transactions. This scatter plot illustrates the relationship between the number of tutor transactions and grade percentages, with a trend line suggesting a positive correlation.}
    \label{fig:unit_1_grades_transactions}
\end{figure}

To supplement this analysis, we built a scatter plot to visualize the association between tutor usage and unit 1 assessment score. The scatter plot in Figure \ref{fig:unit_1_grades_transactions} highlights a positive relationship between the number of tutor transactions and the results of the Unit 1 grade, with a trend line indicating a modest but positive correlation. 
To quantify this relationship, we developed a linear regression model with the number of tutor transactions as the sole predictor variable (see Table \ref{tab:regression_results}). Our results show there is a significant relationship between amount of tutor use and grade. Specifically, the Transactions variable is positively correlated with a higher grade. This suggests that although tutor usage accounts for only a small proportion of the variance in grades, even moderate engagement with the tutor may correspond to improvements in academic performance.
Causal relationships between tutor usage and classroom learning outcomes cannot be inferred directly given the observational nature of this data. However, these results present preliminary evidence of learning as a result of engagement with the tutors.

\section{Discussion}
By analyzing usage metadata from the Apprentice Tutors, academic grade data, and demographic information, this study provides insights into the needs of adult learners engaging with educational technologies. The temporal and demographic data observed in this study improve our understanding of adult-learner interactions with intelligent tutors.

\begin{table}[b!]
\centering
\caption{Regression Analysis Results}
\label{tab:regression_results}
\begin{tabularx}{\columnwidth}{lXXXX}
\hline
\textbf{Variable} & \textbf{Estimate} & \textbf{Error} & \textbf{t-value} & \textbf{p-value} \\
\hline
Intercept & 66.57 & 0.34 & 193.59 & $<$0.001*** \\
Transactions      & 0.26\* & 0.08 & 3.39 & $<$0.001*** \\
\hline
\end{tabularx}
\end{table}

First, adult learners primarily utilized the tutors during evening hours (5 p.m. to 12 a.m.), which aligns with their complex work-life commitments. This pattern suggests the importance of technological flexibility in supporting adult learning and accommodating users based on their unique schedules. In work by \cite{gupta2024intelligent}, learners explicitly mentioned the need for such flexibility.

Second, older learners exhibited greater engagement with the tutors compared to their younger counterparts. We expected that older users would be more hesitant to use the technology and we were surprised to find the opposite.
Several potential explanations may explain this trend. Older adults might have lower baseline comfort with algebra compared to younger students who were more recently exposed to the subject in high school. In addition, differences in intrinsic motivation, patience, or prioritization of learning among older adults could contribute to their increased use of tutors.

Furthermore, tutor usage peaked on Thursdays, which may be strategically linked to the academic schedule, such as examinations typically held on Fridays. This timing allows students to reinforce their understanding and address any uncertainties before assessments, thereby maximizing the tutors' effectiveness in supporting learning outcomes.

Finally, a significant finding emerged from the analysis of the learning curve, demonstrating a reduction in adult learners' error rates with increased participation in practice problems. Although this study is observational in nature and cannot establish causality, the results suggest a positive relationship between tutor usage and skill acquisiton. Tutor users also achieved higher mean grades on the Unit 1 assessments compared to non-users, and our linear regression analysis revealed a statistically significant correlation between the number of tutor transactions (i.e., increased usage) and grades ($p < 0.001$). It is likely that there are some selection effects, where users that choose to use the tutors are also ones that are more likely to do well on the assessment. We would need an experimental manipulation, where we randomly assign users to receive the tutor or not, to determine if tutor usage has a causal relationship with grade. However, by looking only at tutor users and by analyzing learning within-participant in the tutor and between-participant outside the tutor, we believe our findings provide some preliminary evidence that the tutor promotes learning. 

\section{Limitations and Future Work}
An important limitation lies in the non-experimental nature of the study design. Without randomization and a control condition, it is challenging to establish a causal relationship between tutor usage and improvements in academic performance. Although we attempted to isolate the impact of the tutor by focusing on a specific course unit (unit 1), other factors such as prior knowledge and motivation, instructor quality, peer study groups, and access to additional educational resources could have influenced the results. As a result, we cannot definitively attribute improvements in grades to tutor usage alone. To address these limitations, future research should consider implementing a randomized controlled trial design, which would allow for a more rigorous examination of the causal effects of tutor use on academic performance. The current experimental design was chosen to gain buy-in from instructors, as introducing a randomized trial initially could have hindered adoption. Now that we have introduced the tutors, this sets the stage for conducting a randomized trial in the future. 

Despite these limitations, this work provides valuable insight into how adult learners interact with and benefit from intelligent tutors. Future research should adopt more rigorous study designs, such as randomized controlled trials, to more clearly establish causality. Currently, tutor usage among adult learners is optional and supplementary, which presents challenges in systematically controlling usage levels. Although this provides useful data related to user motivation, it may serve us well to closely collaborate with the institution to require tutor use in certain classrooms, while maintaining a control group without tutor access, which would allow more robust comparisons and provide stronger evidence regarding the effectiveness of the tutor.

Finally, current tutor coverage is limited in scope, focusing on a subset of course units and topics. Although we have further expanded tutor coverage since conducting our study, adding more tutors to cover all units within the syllabus would enable a longitudinal assessment of student performance throughout the academic period. Furthermore, applying the tutor approach to subjects beyond mathematics could reveal differences in how learners seek and benefit from supplementary tutors in various disciplines. This broader perspective would enhance our understanding of the conditions under which intelligent tutors are most effective for adult learners at scale.

\section{Conclusion}
This study provides new insight into how adult learners engage and benefit from intelligent tutors in settings where tutor usage is voluntary. By examining both in-tutor performance metrics and classroom assessment outcomes, we have shown that adult learners not only use tutors for flexible, after-hours learning, but also demonstrate skill growth over time as reflected in decreased error rates in the tutors. Moreover, while we cannot assert causality from the Unit 1 observational data, the analysis suggests a positive relationship between tutor usage and improved performance on class assessments.

We aim to design and deploy intelligent tutors that respect the unique constraints and preferences of adult learners, many of whom balance multiple responsibilities and seek educational tools that align with their personal schedules and goals. In addition to performance, we investigate important demographic factors, including age, ethnicity, and gender, that shape engagement. These insights are crucial for developers, educators, and institutions, so they can offer effective supplementary pedagogical technologies for adults.

\section*{Acknowledgments}

This work was generously funded in part from several sources, including the NSF National AI Institutes program (\#2247790 and \#2112532). The views, opinions and/or findings expressed are those of the author and should not be interpreted as representing the official views or policies of the Department of Defense or the U.S. Government. We also thank the members of the Teachable Artificial Intelligence (TAIL) Lab for their feedback and suggestions, as well as Kaitkyn C. Crutcher for invaluable assistance in data analysis and visualization.   

\bibliographystyle{ACM-Reference-Format}
\bibliography{main}

\end{document}